%% file: main.tex
\documentclass[sigconf]{acmart}
\setcopyright{none}
\renewcommand\footnotetextcopyrightpermission[1]{} 
\usepackage{import}
\usepackage{graphicx}
\usepackage{amsmath}
\usepackage{tabularx}
\usepackage{colortbl}
\usepackage{pgfplotstable}
\usepackage{multirow}
\usepackage{color}
\usepackage{multicol}
\usepackage{booktabs}
\usepackage{array}
\usepackage{makecell}
\usepackage{caption}
\usepackage{subcaption}

\newcommand{\etal}{\emph{et al.} }
\newcommand{\methodname}{VizADS-B }

\acmConference[]{Sefi Akerman}{Edan Habler}{Asaf Shabtai}

\begin{document}

\title[VizADS-B: Analyzing Sequences of ADS-B Images Using Explainable Convolutional LSTM Encoder-Decoder]{VizADS-B: Analyzing Sequences of ADS-B Images Using Explainable Convolutional LSTM Encoder-Decoder to Detect Cyber Attacks}

\author{Sefi Akerman, Edan Habler, Asaf Shabtai}
\affiliation{
\institution{Department of Software and Information Systems Engineering, \\ Ben-Gurion University of the Negev}
}

\input{abstract.tex}


\maketitle

\input{introduction.tex}
\input{adsb.tex}
\input{adversary.tex}
\input{relatedworks.tex}
\input{method.tex}
\input{evaluation.tex}

\input{conclusions.tex}

\bibliographystyle{ACM-Reference-Format}
\bibliography{main.bib}

\end{document}

%% file: abstract.tex
\begin{abstract}
The purpose of the automatic dependent surveillance broadcast (ADS-B) technology is to serve as a replacement for the current radar-based, air traffic control (ATC) systems.
Despite the considerable time and resources devoted to designing and developing the system, the ADS-B is well known for its lack of security mechanisms.
Attempts to address these security vulnerabilities have been made in previous studies by modifying the protocol's current architecture (e.g., adding encryption or authentication mechanisms) or by using additional hardware components (e.g., sensors for verifying the location of the aircraft by analyzing the physical signal).
These solutions, however, are considered impractical because of 1) the complex regulatory process involving avionic systems, 2) the high costs of using hardware components, and 3) the fact that the ADS-B system itself is already deployed in most aircraft and ground stations around the world.
In this paper, we propose \methodname, an alternative software-based security solution for detecting anomalous ADS-B messages, which does not require any alteration of the current ADS-B architecture or the addition of sensors.
According to the proposed method, the information obtained from all aircraft within a specific geographical area is aggregated and represented as a stream of images.
Then, a convolutional LSTM (ConvLSTM) encoder-decoder model is used for analyzing and detecting anomalies in the sequences of images.
In addition, we propose an explainability technique, designed specifically for convolutional LSTM encoder-decoder models, which is used for providing operative information to the pilot as a visual indicator of a detected anomaly, thus allowing the pilot to make relevant decisions.
We evaluated our proposed method on five datasets by injecting and subsequently identifying five different attacks.
Our experiments demonstrate that most of the attacks can be detected based on spatio-temporal anomaly detection approach.
\end{abstract}

%% file: introduction.tex
\section{\label{sec:intro}Introduction}

Currently, air traffic management system consists of two main subsystems, the primary surveillance radar (PSR) and the secondary surveillance radar (SSR).
Both systems enable measuring the aircraft's range and bearing from radar stations.
In recent years, the number of commercial flights has significantly increased\footnote{\url{https://www.iata.org/publications/Documents/iata-annual-review-2018.pdf}} and as a result, the shortcomings of traditional air traffic management systems have become more apparent and concerning, especially over oceans or mountainous areas where it is difficult to employ radar systems, or in very crowded areas such as airports.
In this context, the aviation community is actively taking steps to move from uncooperative and independent air traffic surveillance (e.g., PSR and SSR) to cooperative dependent surveillance systems (CDS). 
The purpose of this shift is to substantially reduce air traffic control (ATC) costs and, at the same, time optimize the navigation process, thus improving flight safety.

Automatic dependent surveillance-broadcast (ADS-B)~\cite{blythe2011ads} is an implementation of a CDS system certified by the Federal Aviation Administration (FAA) and the International Civil Aviation Organization (ICAO), which has already been deployed in aircraft and ground stations across Europe, Canada, and Australia.
The FAA has decided to use the ADS-B system for all aircraft movement in the US by 2020.
ADS-B uses GPS navigation to provide more accurate information regarding the aircraft state.
This aircraft information is available to ground stations and ATCs, as well as the aircraft themselves, which can use the information to derive and share their position and flight direction using the ADS-B transponder (transmitter-responder).

Despite the great importance of the system in the near future, it is evident that most of the effort invested in planning and developing the ADS-B system was placed on accuracy and cost savings, at the expense of security which was not considered.
Pushing security to the sidelines has resulted in fundamental flaws in the current system. 
In particular: 1) messages are not broadcast with an authentication code or digital signature and therefore can be replayed, manipulated, or forged, 2) messages are broadcast in simple, publicly known formats, leaving them exposed to eavesdropping, and 3) authorized aircraft or ATC stations don't have to authenticate before transmitting; thus, there is no way to distinguish between authorized and unauthorized entities.
As a case in point, in recent years, researchers in academia and industry have pointed out weaknesses in the system's design and implementation, and demonstrated how the system can be tampered with using off the shelf hardware and software~\cite{schafer2013experimental,costin2012ghost}.

In order to address the risks that have been identified, recent research has suggested the use of encryption~\cite{finke2013enhancing} to prevent eavesdropping; aircraft authentication via challenge-response~\cite{kacem2015key}; message authentication~\cite{costin2012ghost,feng2010data} in order to provide secured message broadcasting; and validation of the data transmitted from the planes by adding additional sensors in space.
However, implementing these solutions at this stage is impractical because the ADS-B system is already deployed in most aircraft.
In addition, due to the strict regulation process regarding the implementation of avionic systems, applying such modifications to the current protocol requires a great deal of resources and a significant amount of time to implement and deploy (for that matter, the ADS-B protocol design and development started in the early 1990s).

In order to avoid the challenges and high costs associated with hardware and software changes, Habler \etal~\cite{habler2018using} proposed a machine learning based approach for detecting manipulated ADS-B messages sent by an attacker or compromised aircraft. 
The authors utilized an LSTM encoder-decoder algorithm for modeling flight routes by analyzing sequences of legitimate ADS-B messages, applied these models to evaluate ADS-B messages received and identify deviations from the legitimate flight path.
This approach, however, analyzes the data of each aircraft individually, and it does not consider the spatial-temporal correlation between aircraft sharing a common airspace.

In this research, we follow the approach of Habler \etal~\cite{habler2018using} and present \methodname, an extended software-based security solution for detecting anomalous ADS-B messages that does not require alteration of the current ADS-B system.
\methodname represents multiple ADS-B messages (received from different aircraft) within a geographic area as a stream of visual images, and applies a convolutional LSTM (ConvLSTM) encoder-decoder model to detect anomalous images, thus providing an indication of anomalous ADS-B messages. 
By using this representation approach, we can process information obtained from multiple aircraft, \textit{without the need for the complex process associated with feature engineering and extraction} which is a challenging task, especially in dynamic real-time use cases. 
In addition, \methodname applies an \text{explainability technique}, designed specifically for convolutional LSTM encoder-decoder models.
The explainability function is used for providing operative and supporting information to the pilot in the form of a visual indicator of the detected anomalies.
These visual indicators make it easier to the pilot in making correct decisions as well as in quickly identifying false alarms.

As mentioned by Habler \etal~\cite{habler2018using}, since such an anomaly detection approach only requires receiving feeds of ADS-B messages (provided by the ADS-B system) and visualizing classification outputs and anomalies, it can be implemented as a dedicated application installed on the electronic flight bag (EFB) server and does not require any change to the ADS-B system itself.

We evaluated VizADS-B using five datasets, each of which contains flights from a 10,000 square kilometer area, centered around different airports, which were recorded for a period of seven days.
Our experiment included the injection of five different types of anomalies and we showed that our approach was able to successfully and clearly identify most of the attacks.

To summarize, our contributions in this study are as follows.

\begin{itemize}
    
    \item we show that the data of multiple aircraft sharing and flying within the same airspace can be represented by a sequence of images and thus can be modeled and analyzed by convolutional recurrent neural networks;
    
    \item we show that the ConvLSTM encoder-decoder model can be used to amplify the anomaly scores of malicious ADS-B messages, allowing the identification of anomalous images; 
    
    \item our approach does not require changes to the architecture of the ADS-B protocol or additional sensors;
    
    \item we introduce a new explainability technique, which, to the best of our knowledge, is the first to explain the output of a convolutional encoder-decoder model;
    
    \item we propose using an explainability technique in the detection of attacks in operational, real-time, cyber-physical systems, as well as to effectively handle false alarms;
    
    \item our proposed approach is adaptive and flexible, and therefore it can be applied to additional geographical areas without the need for specific modifications.
\end{itemize}

%% file: adsb.tex
\section{\label{sec:adsb}The ADS-B System}

The FAA began a process of updating and modernizing the national airspace systems (NAS), known as the next generation air transportation system (NextGen). 
NextGen's goals are to increase flight traffic safety, capacity and flexibility, as well as to reduce dependency on outdated radar infrastructure. 
The ADS-B system is a key component of NextGen.
ADS-B is a satellite-based, `radar-like' system designed to continuously derive the aircraft position from the global navigation satellite (GPS) system. 
ADS-B provides the aircraft velocity, identity, and position with greater accuracy, providing a clearer picture of the air traffic. 

\begin{figure}[h]
\centering
\includegraphics[width=9cm,keepaspectratio]{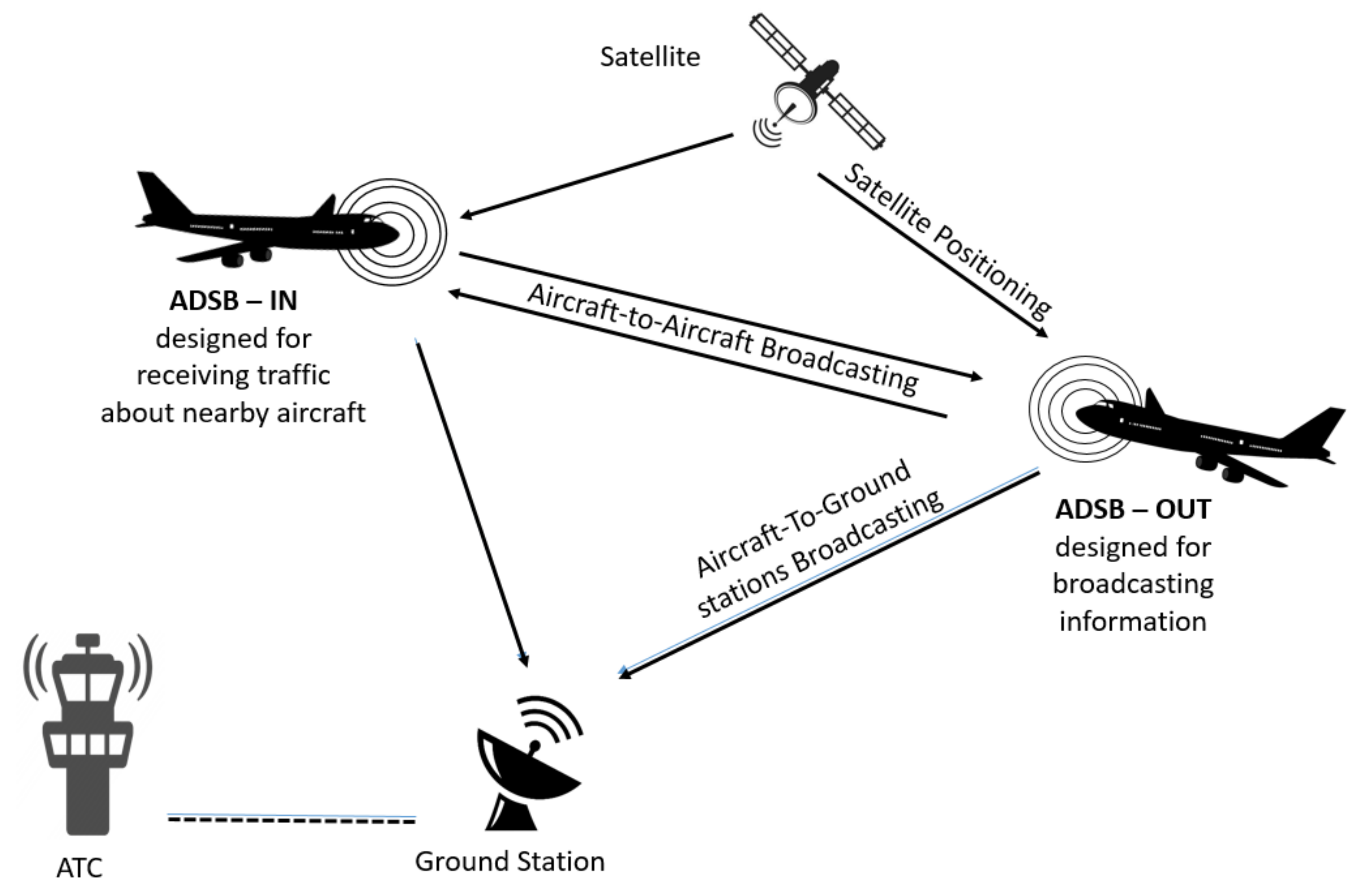}
\caption{Illustration of the ADS-B protocol components.
Using the ADS-B IN component, the aircraft get their position from satellite system. 
The ADS-B OUT component is used to broadcast related information. 
Nearby aircraft and ground stations receive these messages with the ADS-B IN component. 
The information that achieved by the ground stations is then transmitted to the ATC.}
\label{fig:ADSB-structure}
\end{figure}

ADS-B includes two separate systems: ADS-B In and ADS-B Out.
The ADS-B In system allows an aircraft to receive and display messages transmitted by another aircraft within the receiving range. 
The ADS-B Out system allows an aircraft to continuously generate and broadcast messages over unencrypted VHF or satellite-based data link networks.
The transmitted messages generated by the ADS-B Out system are then processed by nearby aircraft and ATC stations located on the ground.
This process is illustrated in Figure~\ref{fig:ADSB-structure}.

%% file: adversary.tex
\section{\label{sec:adversary}Adversary Model}

The ADS-B protocol was originally intended to increase message availability among aircraft and ground stations.
As a result, the protocol does not include security mechanisms and the messages are broadcast unencrypted; thus, anyone with a compatible receiver can obtain the data contained in the messages.
While in the late 1980's it may have been reasonable to assume that access to such hardware is not trivial, it is currently not the case when these components can easly be purchased at a low price.
This distinction is essential for understanding the plausibility of disrupting the ADS-B system.
In the same manner as obtaining ADS-B message passively, commercial off-the-shelf (COTS) products can be used to inject ADS-B messages into the air space in order to perform a wide range of attacks such as spoofing and flooding. 

It is reasonable to assume that an attacker will have the means and ability to both eavesdrop and obtain the range and inject false data, with the use of  a software-defined radio (SDR) transceiver.
Moreover, it is also plausible to assume that in order to maintain secrecy, an attacker will follow the constraints that apply to all of the aircraft. 
In order to identify attackers who use COTS products, previous work has suggested the use of different techniques in order to compare measurements between several sensors located in different ground stations, e.g., the use of the multilateration (MLAT) technique, which is based on measurement of the time of arrival (TOAs) of energy waves between a varying number of points.
However, there are ways to make it difficult for these signal analysis detection mechanisms, such as using a UAVs or drones, in addition, an attacker may be on the aircraft itself (as described in~\cite{mirzaei2019security}).

\begin{figure}[h!]
\centering
\includegraphics[width=0.5\textwidth,keepaspectratio]{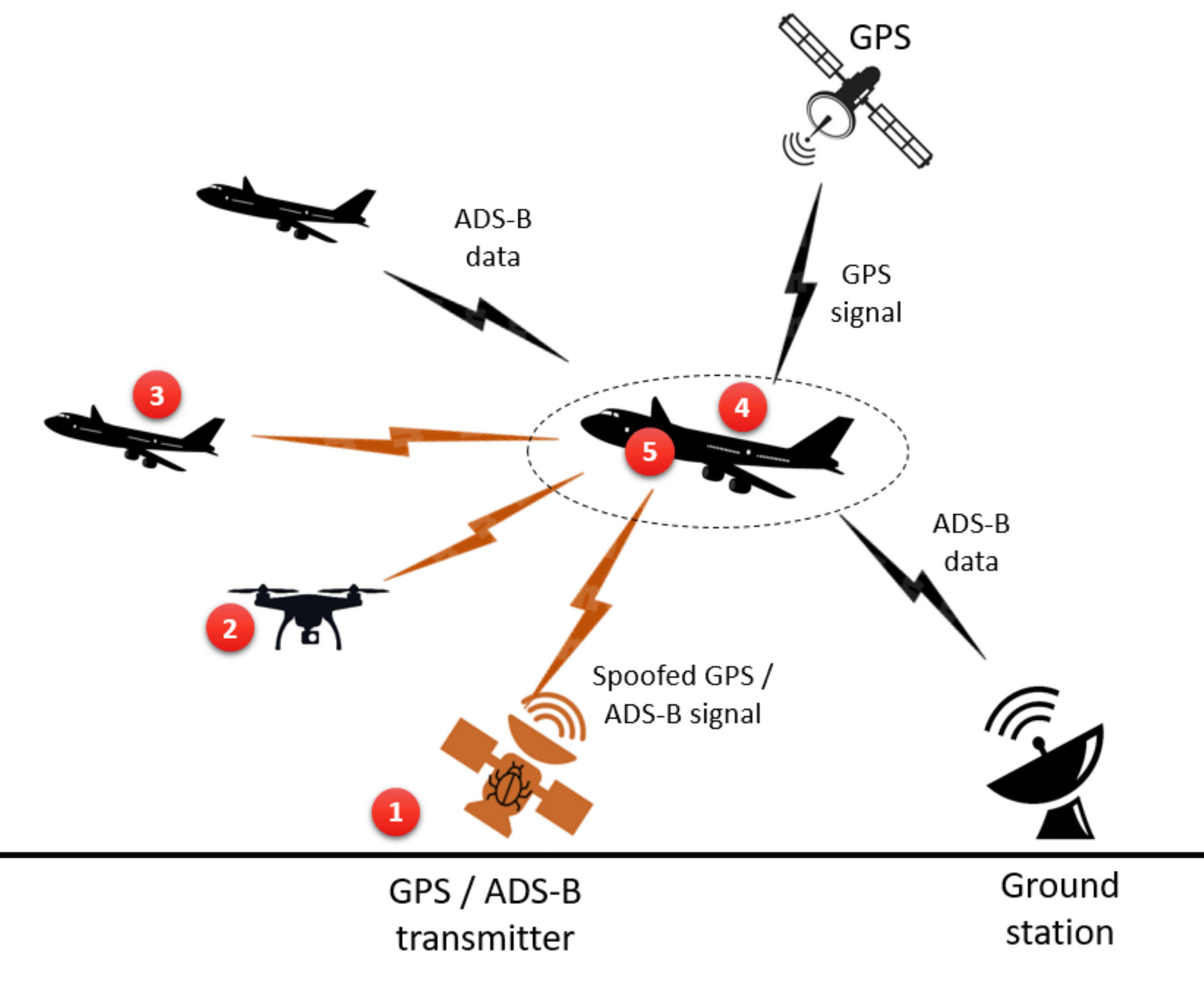}
\caption{Visual illustration of potential threats.}
\label{fig:ADSB-Vector}
\end{figure}

The potential hostile participants and threats to the ADS-B system are illustrated in Figure~\ref{fig:ADSB-Vector}).
The first and most obvious participant is an attacker located on the ground who uses an ADS-B receiver and transmitter to disrupt or affect the air space.
An attacker with more resources and capabilities may even use drones and UAVs (as illustrated in labels two and three respectively), which may improve the operating state of the attacker, while challenging security mechanisms that take into account the TOAs of energy waves.

Beyond the external participant affecting the air space, one must consider an attacker that is, in one form or another, on board the aircraft itself (as illustrated in labels four and five). 
Such a passenger is able to influence the aircraft's receiver by transmitting false messages; in addition, the passenger can mislead nearby aircraft by transmitting false messages, and these messages may also conflict with legitimate ADS-B messages transmitted by the aircraft on which the attacker is located. 
These actions and others can be performed by a hostile takeover of the aircraft's systems by means of a cyber attack aimed at affecting the aircraft's receivers and transmitters or the pilots flight instruments.

%% file: relatedworks.tex
\section{Related Work}

\subsection{Security of the ADS-B System}

Previous research has investigated the security challenges associated with the ADS-B system and proposed various methods and solutions for protecting the system. 
The main ideas in previous studies (summarized in Table~\ref{tab:relatedworks}) include encryption, physical layer analysis, and a multilateration technique. 
Recently, a new machine learning approach of using machine learning for detecting anomalous ADS-B messages was suggested.

\newcolumntype{C}[1]{>{\centering\let\newline\\\arraybackslash\hspace{0pt}}m{#1}}

\begin{table*}
\centering
\caption{Summary of related work.}
\begin{tabular}{ | C{3.5cm} | C{5.5cm} | C{5.5cm} |} 
 \hline
 \textbf{Approach (references)} & \textbf{Protocol modification required} & \textbf{Additional sensors / entities required} \\ 
 \Xhline{3\arrayrulewidth}

 \rowcolor[gray]{0.9}
 {Encryption~\cite{costin2012ghost,finke2013enhancing,feng2010data,strohmeier2015security}} & {Requires an encryption mechanism and a new type of message for key exchange.} & {Key exchange requires a trusted entity among the aircraft.} \\
 \hline
 
 {Doppler effect and PHY-layer analysis~\cite{ghose2015verifying,costin2012ghost}} & {Relies on available information from the physical layer and the pulses emitted from the transponders, therefore no protocol modification is required.} & {Measurements rely upon ground stations or other entities.} \\ 
 \hline

\rowcolor[gray]{0.9}
 {Multilateration \& group verification~\cite{finke2013enhancing,strohmeier2015lightweight,smith2006methods}} & {Verification via a group of users requires a new message type and protocol in order to establish a trust mechanism.} & {These methods are based on measuring the signal arrival time from multiple points and therefore, require more than one sensor or entity.} \\ 
 \hline

{Machine learning based anomaly detection~\cite{habler2018using,sun2016large,strohmeier2018k,zhang2019ensemble}} & {No modifications are needed.} & {No additional sensors or entities are required.} \\ 
 \hline

\label{tab:relatedworks}
\end{tabular}
\end{table*}

\subsubsection{Encryption}
\indent Cryptographic measures have been considered for securing communication in wireless networks. 
Strohmeier \etal~\cite{strohmeier2015security} discussed the question of whether the current implementation of ADS-B can be encrypted, and Finke \etal~\cite{finke2013enhancing} introduced a number of encryption schemes. 
However, the worldwide deployment of the ADS-B system makes encryption key management a challenge. 
Costin \etal~\cite{costin2012ghost} and Feng \etal~\cite{feng2010data} suggested PKI (public key infrastructure) solutions based on transmitting signatures and elliptic-curve cryptography, respectively. 
Another solution discussed in the literature is the use of the retroactive key publication technique, such as the µTESLA protocol~\cite{strohmeier2015security}.
This technique, however, requires modifications to the current mechanism of the protocol by adding a new message type for key publishing.

\subsubsection{Physical Layer Analysis and Doppler Effect}
\indent Spoofed message injection attacks are one of the most dangerous types of attack. 
Strohmeier \etal~\cite{strohmeier2015intrusion} proposed an intrusion detection system, based on physical layer information and a single receiver, in order to detect such attacks on critical air traffic infrastructures without the need for additional cooperation of the aircraft. 
Another solution suggested by Ghose \etal~\cite{ghose2015verifying} is to verify the velocity and position of the aircraft by exploiting the short coherence time of the wireless channel and the Doppler spread phenomenon. 
A method presented by Schafer~\cite{schafer2016secure} is based on verification of the aircraft motion using the Doppler effect. 
Both options rely upon the participation of ground stations or other entities.

\subsubsection{Multilateration and Group Verification}
\indent In the last decade, signal analysis has been successfully employed in the fields of wireless communication. 
One popular form of cooperative independent surveillance that has been used in military and civil applications is multilateration (MLAT). 
MLAT is a navigation technique based on the measurement of the time difference of arrival (TDOA) between at least two stations at known locations. 
Smith~\cite{smith2006methods} provided a method based on MLAT that can serve as a backup for ADS-B communication. 
Additional work based on MLAT suggested by Schafer \etal~\cite{schafer2015secure} showed that it is possible to verify a 3D route with a group of four verifiers. 
In addition, Strohmeier \etal~\cite{strohmeier2015lightweight} suggested a method of continuous location verification by computing the difference in the expected TDOA between at least two sensors.

\subsubsection{Machine Learning}
Machine learning models are used in many domains for detecting anomalies in data streams.
In the context of aviation, the large scale of the data can be used to solve tasks and learn patterns using machine learning, as suggested by Junzi \etal~\cite{sun2016large}. 
Several studies involving machine learning, have already been conducted in the aviation field. 
Stromeier \etal~\cite{strohmeier2018k} used the kNN algorithm which succeeded in outperforming traditional methods for the localization of aircraft. 
Zhang \etal~\cite{zhang2019ensemble} presented an ensemble of models to identifying aircraft accidents.
Strohmeier \etal~\cite{strohmeier2018real} introduced the possibility of using machine learning to exploit the lack of privacy for the purpose of inferring confidential corporate mergers and government relations from broadcast ADS-B messages.
In addition, Habler \etal~\cite{habler2018using} presented a method that can characterize the nature of aircraft movement.
As mentioned, an advantage of this method is that there is no need for additional sensors or architectural changes.

\subsubsection{Summary of Related Work}
Each of the methods described above has its advantages and disadvantages.
Adding encryption measures to the ADS-B system provides confidentiality and the ability to authenticate  messages, thus protecting against spoofing and eavesdropping attacks.
However, previously proposed solutions that use this technique \cite{costin2012ghost,feng2010data,finke2013enhancing,strohmeier2015security} require redesigning and implementing the protocol (i.e., a new type of message and key exchange scheme). 
As described above, these requirements involve protocol modifications and adding a trust mechanism between the aircraft. 
Due to the regulation process regarding avionic systems and the fact that the ADS-B system is already deployed in most aircraft, applying such modifications to the current protocol at this stage would require a great deal of resources and a significant amount of time and funds to implement and deploy.

Approaches that rely upon the Doppler effect and multilateration (\cite{ghose2015verifying,schafer2016secure}) are able to provide accurate location verification, thus making it possible to detect spoofing attacks. 
However, suggested solutions using these approaches rely on the presence of sensors on the ground, which isn't a given in some geographic areas. 
In addition, proposed solutions utilizing these approaches also require protocol modifications in order to establish the trust mechanism between airplanes, as well as to avoid reliance on sensors located on the ground~\cite{schafer2015secure}.
Furthermore, \etal~\cite{shang2019multidevice} demonstrated that a sophisticated attacker can manipulate multilateration based solutions by using several broadcasting sources.

Habler \etal~\cite{habler2018using} proposed an alternative security solution, which differs from the solutions suggested in the above-mentioned works in that it interacts with the existing system, and thus does not require any modification or additional hardware components. 
This method is based on a machine learning technique which analyzes information from flight routes and the behavior of airplanes on these routes.
A dedicated model is trained for each flight route, and therefore various information is needed to create the various models. 
Since the work specified above is done for each flight route separately, our proposed \methodname method can be extended in future work by integrating the anomaly score derived according to the method proposed by Habler \etal~\cite{habler2018using}.
This can be done by expressing the individual anomaly scores derived from their method within the images analyzed by \methodname.

\subsection{\label{subsec:anomalydetection}Anomaly Detection in a Stream of Images}

Machine learning algorithms are used in order to solve complex problems that are not easily solved in traditional ways.
Convolutional neural networks (CNNs) have been shown to be effective at image analysis, particularly in identifying anomalies within images~\cite{gavrilut2007integrated,sampaio2011detection}.
Their representation and processing capabilities can reduce the need for the feature engineering process, since the CNN can detect and extract complex patterns.

Several attempts have been made in order to leverage the advantages of traditional CNNs for solving problems involving sequences of images. 
Most of these attempts involved some combination of recurrent neural networks (RNNs) and CNNs, and a few architecture types based on  various combinations of the two were used~\cite{kravchik2018detecting}.
In the domain of spatio-temporal anomaly detection, several studies used an architecture consisting of layers that combine the long-short term memory cells (LSTM) with CNNs~\cite{xingjian2015convolutional,medel2016anomaly} and showed that this combination performed well.
In this research we opted to use the Conv-LSTM method proposed by Shi \etal~\cite{xingjian2015convolutional}, because it is considered a state-of-the-art method particularly suited to our interest in the detection of spatio-temporal anomalies within a sequence of images.

\subsection{Explainable AI}
Explainable AI (XAI) is an emerging researched field in machine learning whose purpose is to allow users to understand, confidently trust, and effectively manage the next generation of AI products~\cite{gunning2017explainable}.
Most of the methods developed in recent years are meant to explain supervised machine learning models. 
For example, the LIME~\cite{ribeiro2016should} method introduced for explaining the prediction using a local mode; the DeepLIFT method~\cite{shrikumar2017learning}, which uses back propagation through all of the neurons in the network to explain the output; and the SHAP~\cite{lundberg2017unified} method, which is a unified approach that aims to explain the model output using shaply value and game theory properties.

The need to explain the output is especially important in anomaly detection based on deep learning models, because usually in this case not all of the anomaly types are known (labeled).
Providing an explanation of a detected anomaly will enable the human expert to trust the model and make the correct decisions.
An initial attempt at explaining an autoencoder was made by Antwarg \etal~\cite{antwarg2019explaining}, and proposed using SHAP to explain the output of an RNN autoenocder.
Another relevant study performed by Arnaldo \etal~\cite{arnaldo2019ex2}, who tried to detect outliers using a PCA model and explain its output by displaying the most significant features. 
Similarly, Nguyen \etal~\cite{nguyen2019gee} offered a technique for explaining variational autoencoders by finding the most significant features based on their gradient values.
In this study, we propose a method for explaining the output of a convolutional LSTM autoencoder.

%% file: method.tex
\section{\label{sec:method}Proposed Method: \methodname}

\subsection{Motivation}
The ADS-B system allows the pilot to view information about the surrounding aircraft by receiving messages that are broadcast by these aircraft and creating an updated visual state of its surrounding area.
Because of the known security issues of the ADS-B system, and specifically since it is possible to manipulate or broadcast forged messages, the ability to detect malicious ADS-B messages is crucial.  

Detecting anomalies in data streams received by several aircraft simultaneously is a challenging task, since the sequences of messages sent by individual aircraft must be examined, in addition to the relationship (correlation) between messages sent by different aircraft in the same geo-location and period of time which also must be examined.

In order to facilitate the representation and analysis of multiple ADS-B messages broadcast by different aircraft, we opt to represent the information obtained from the broadcast ADS-B messages in the form of a stream of images. 
This approach eliminates the need for feature engineering, which can be complicated in this scenario (for example, because of the unknown number of aircraft in a given location at any given point in time).

In order to analyze the data we use an convolutional LSTM encoder-decoder network~\cite{finn2016unsupervised} to detect spatio-temporal anomalies in sequences of images generated from the data streams of ADS-B messages received.

\subsection{Constructing the Images}
We aim to include all of the data contained in an ADS-B message and represent the relative proximity among nearby aircraft within an image, thus enabling the model to learn meaningful complex patterns from the data.
An ADS-B message contains the aircraft's altitude, latitude, longitude, speed and heading direction, as presented in Table~\ref{tab:messageinfo}.

\begin{table*}
\centering
\caption{\label{tab:messageinfo}ADS-B message attributes.}

\begin{tabular}{ | C{1.6cm} | C{4.5cm} | C{1cm} | C{2cm} | C{5.5cm} |} 
 \hline
 \textbf{Attribute} & \textbf{Description} & \textbf{Unit} & \textbf{Value range} & \textbf{Represented in image} \\ 
 \Xhline{3\arrayrulewidth}

 \rowcolor[gray]{0.9}
 {Longitude} & {aircraft's east-west position on the Earth's surface} & {deg} & {[-90, 90]} & {by the arrow's position} \\
 \hline
 
 {Latitude} & {aircraft's north-south position on the Earth's surface}  & {deg} & {[-180, 180]} & {by the arrow's position} \\
 \hline       
 
 \rowcolor[gray]{0.9}
 {Speed} & {aircraft's speed relative to the ground} & {Knt} & {[0, 500]} & {by the length of the arrow, and the difference in the arrow's location between consecutive images} \\
 \hline
 
 {Altitude} & {aircraft's altitude above sea level} & {ft} & {[0, 40K]} & {by the arrow's size} \\
 \hline
        
 \rowcolor[gray]{0.9}
  {Heading} & {aircraft's direction of progress} & {deg} & {[0, 360]} & {by the arrow's direction} \\
 \hline
 
 {Time} & {epoch standard time} & {ms} & {varies} & {by the image in the sequence} \\
 \hline
 
 \rowcolor[gray]{0.9}
 {Call-sign} & {aircraft's unique identifier} & {-} &  {-} & {by a unique arrow within the image} \\
  \hline

\end{tabular}
\end{table*}

\begin{figure}[!h]
\centering
\includegraphics[width=8cm,keepaspectratio]{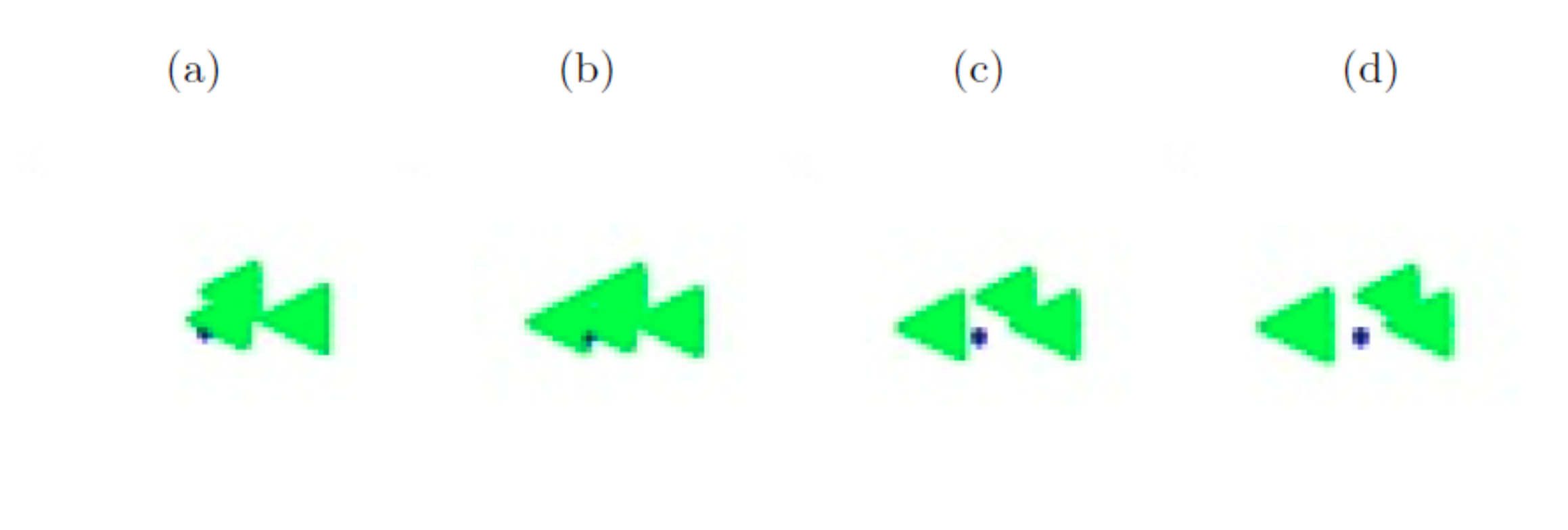}
\caption{An example of a sequence of images representing the ADS-B messages in consecutive time frames (a-d).} 
\label{fig:seq-of-images}
\end{figure}

Assume an entity $e$ (an aircraft or ground station) that receives ADS-B messages broadcast by nearby aircraft \{$a_1$,..., $a_n$\}. 
In order to apply the proposed method, the messages received by $e$ are transformed into a sequence of $s$ images (see Figure~\ref{fig:seq-of-images}).
Each image represents a predefined time interval $\Delta t$ and summarizes the information reported - using the ADS-B protocol - by all aircraft within a specific geographical area at $\Delta t$. 
As shown in Figure~\ref{fig:seq-of-images}, 
each aircraft is represented by an arrow that encapsulates the location, speed, direction and altitude of the aircraft. 
In addition, in cases in which an anomaly score is provided for each ADS-B message individually, the anomaly score is represented by the color of the arrow. 

\begin{figure*}[!t]
\centering
\includegraphics[width=16cm,keepaspectratio]{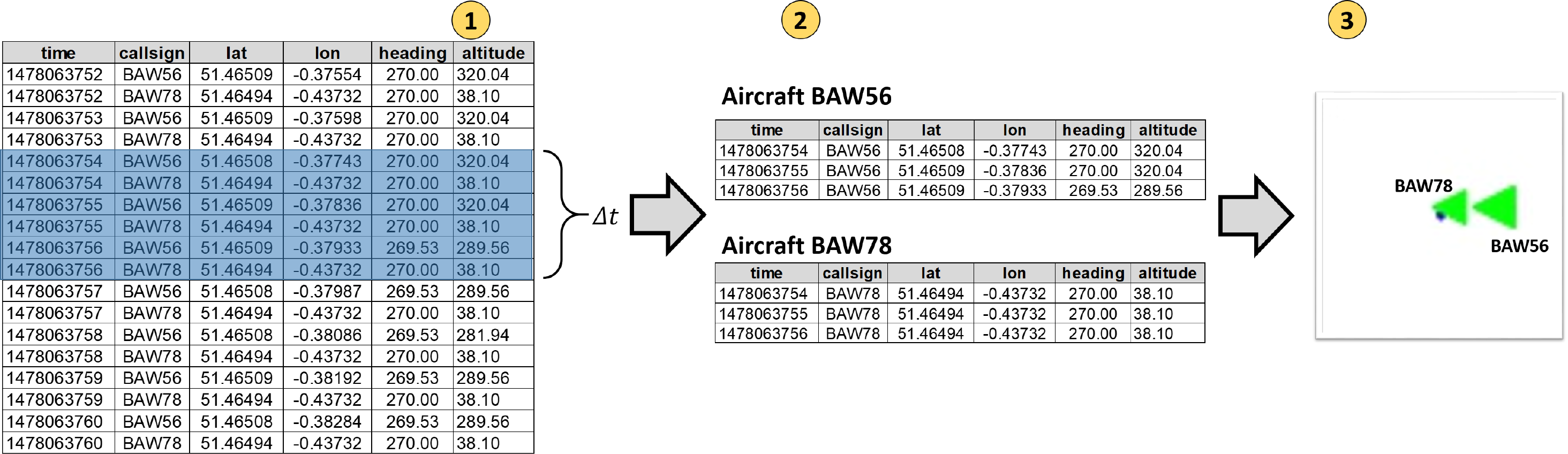}
\caption{An example of the image creation process.}
\label{fig:messages2image}
\end{figure*}

The images are generated as follows. 
Given the geographic location (altitude, latitude, longitude, and heading direction) of all aircraft in the selected area, we first perform a projection to a 2D plane using Lambert conformal conic projection (LCC)~\cite{snyder1982map}.
Then, given a sequence of ADS-B messages $\{m_1,m_2, ..., m_k \}$ sent by $a_i$ within $\Delta t$, we create an arrow representing the following properties: 
The start point of the arrow is set to be $(m_1(latitude),m_1(longitude))$ and the end of the arrow is set to be $(m_k(latitude),m_k(longitude))$, both indicating the position of the aircraft in the air. 
Consequently, the length of the arrow represents the distance traveled during $\Delta t$.
The arrow's direction is calculated by:
\[ f_{direction}(m_1(heading) ,...,m_k(heading)) \]
The size of the arrow represents the altitude of the aircraft: \[ f_{altitude}(m_1(altitude) ,...,m_k(altitude)) \] 
Finally, the color of the arrow represents the anomaly score of the messages and is calculated by: \[f_{anomaly}(m_1(anomalyScore) ,...,m_k(anomalyScore)) \]
In our case, we used the identity function for $f_{direction}=m_k(heading)$ and a logarithmic relation for $f_{altitude}=log(m_k(altitude))$.
An example of this process is illustrated in Figure~\ref{fig:messages2image}.
All ADS-B messages in $\Delta t$ are selected (step 1) and aggregated for each individual aircraft (step 2).
Then, the defined functions are used to transform the data into an image (step 3).
Repeating this process for each point in time results in a stream of images that contains meaningful information regarding all of the aircraft in the area.

\subsection{Model Description}
\subsubsection{Overview of Convolutional LSTM Encoder-Decoder.}

\begin{figure}
\centering
\includegraphics[width=9cm,height=10cm,keepaspectratio]{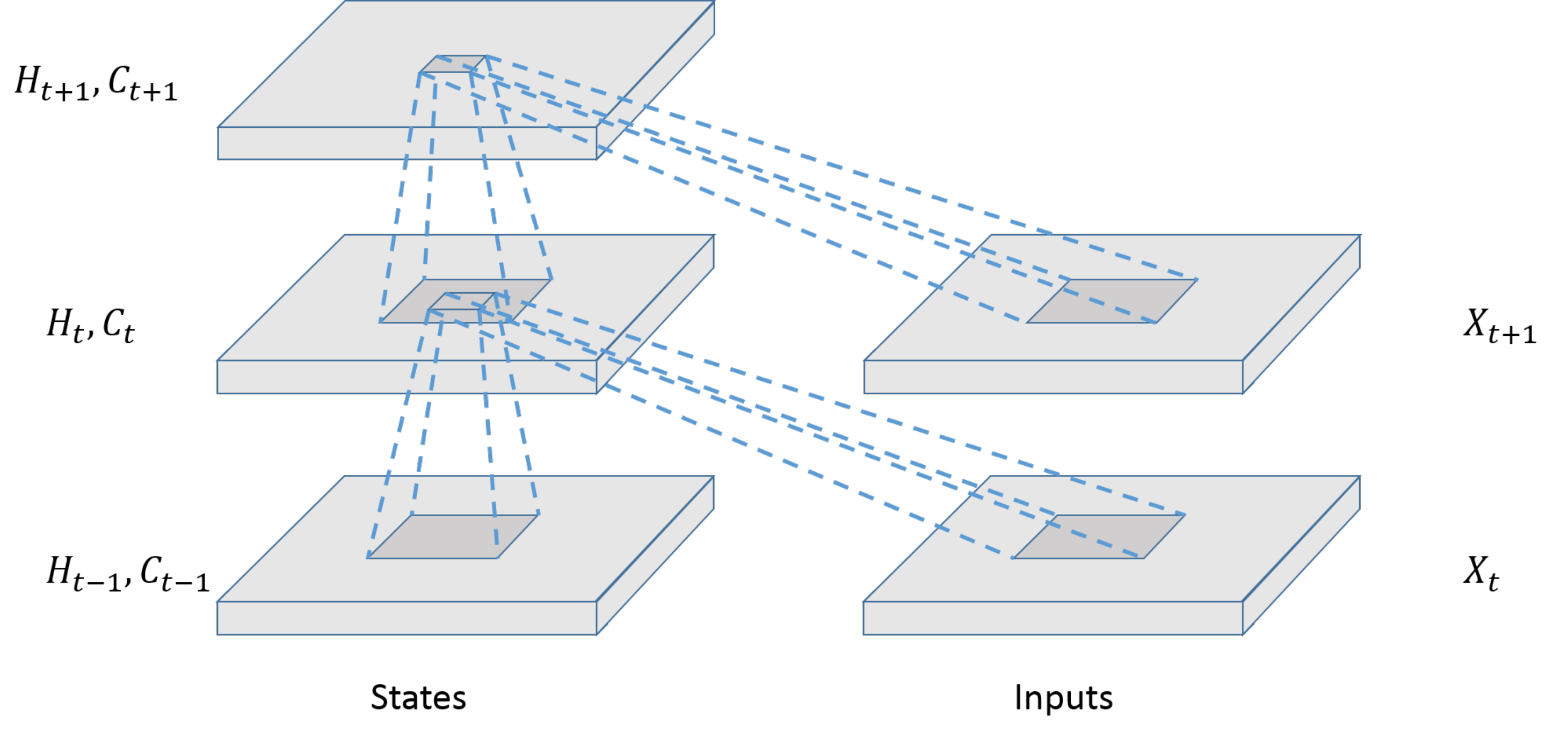}
\caption{Conv-LSTM layer structure illustration.
$X$ denotes the input, $H$ denotes the hidden states, and $C$ denotes the current states.}
\label{fig:convlstm}
\end{figure}

As described in Section~\ref{subsec:anomalydetection}, the CNN neural network architecture has been shown to work very well in many cases, primarily in cases related to image processing~\cite{gavrilut2007integrated,sampaio2011detection}. 
A CNN based on long short-term memory (LSTM) is designed to deal with spatio-temporal data such as videos.
First introduced by Shi \etal~\cite{xingjian2015convolutional}, it has been since proven to perform well in several subsequent studies~\cite{medel2016anomaly}.
Figure~\ref{fig:convlstm} illustrates one layer of a Conv-LSTM network, where $X$ denotes the input, $H$ denotes the hidden states, and $C$ denotes the current states at different time periods (represented by $t$).

Our model's architecture consists of two main components: encoder and decoder. 
The encoder component involves LSTM- convolutional neural network layers which are used for feature extraction and sequence prediction. 
The decoder component is also composed of LSTM-convolutional neural network layers and is used to reconstruct the input received by the encoder. 

By applying both encoder-decoder layers as a single unit, we attempt to maximize the probability of capturing the meaningful structure of the input derived from legitimate unlabeled data. 
Therefore, when malicious data is received as input by the decoder component, it will not propagate properly through the model, thus leading to a significant difference between the input and the output which can be used for anomaly detection (i.e., anomalous input).

Using the explanability technique described in Section~\ref{subsec:explainabilty}, we can further analyze the output of the model and provide a human observer (i.e., the pilot) additional visual information that supports real-time decision-making resulting in improving detection rate.

\subsubsection{Training the Model.}
\begin{figure}
\centering
\includegraphics[width=5cm,keepaspectratio]{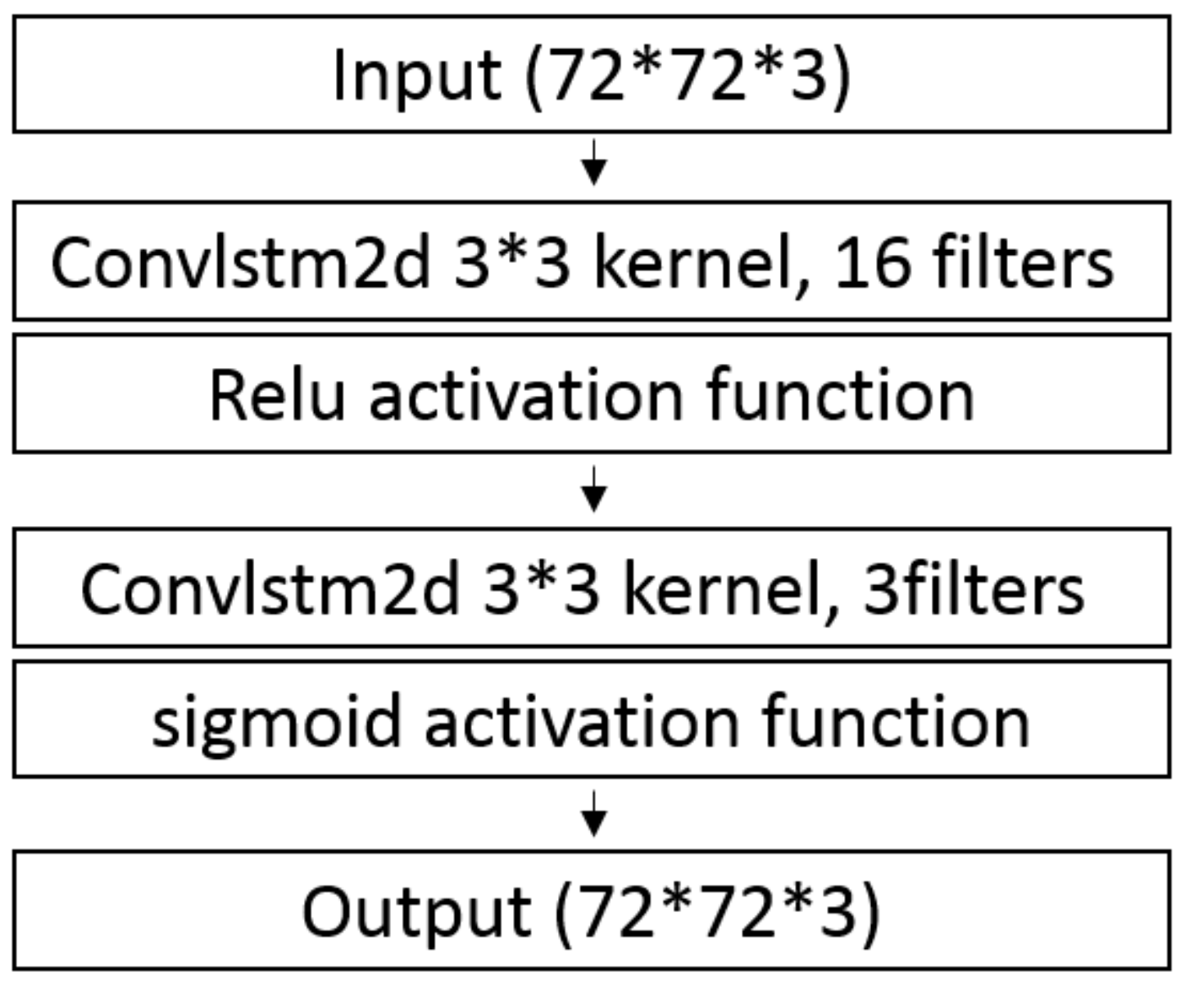}
\caption{Conv-LSTM encoder-decoder structure.}
\label{fig:net_structure}
\end{figure}

In order to reconstruct frames of images with minimal error, we train our model on a training dataset using the Adam optimizer~\cite{wang2004image} and the mean square error (MSE) loss function. 
The structure of the proposed model is displayed in Figure~\ref{fig:net_structure} and consists of shallow encoder and decoder layers with 16 3*3 sized kernels. 
The encoder component learns from sequences of images of length $s$ (denoted as frames) by optimizing its hidden layers to capture meaningful features of the data.
Afterward, the decoder component reconstructs each frame (i.e., a sequence of $s$ images) using the current hidden state of the decoder, the values predicted for the images within the previous frame, and the values of the images within the current frame, similar to RNNs.

\subsubsection{Anomaly detection method.}
After training a model that can reconstruct images that represent legitimate sequences of images (frames), we expect the input and output frames to look alike. 

In order to compare the model's input and output frames, we opt to use the SSIM function~\cite{kingma2014adam}. 
The SSIM index is designed to capture structural similarities between two images rather than pixel-wise similarity, which is less informative when the aim is to evaluate how the model perceives the entire image.

We expect the model to fail in reconstructing images generated from malicious messages.
Consequently, in such cases the SSIM value computed for the model's input and output should be low, thus indicating an anomaly. 

In order to reduce the false alarm rate, we aggregate the anomaly scores computed for $w$ consecutive frames of size $s$. 
We calculate the anomaly score (i.e., SSIM value) for each frame. 
A frame is considered ``suspicious'' if the SSIM value is below a threshold $t_1$ (derived by using a validation set).
We then count the number of ``suspicious'' frames within the window, and if the sum exceeds a second threshold $t_2$, we classify the window as anomalous.

\subsection{\label{subsec:explainabilty}Proposed Explainability Method} 

We propose a simple and yet efficient explanability technique, which, to the best of our knowledge, is the first to explain the output of a convolutional encoder-decoder models.
The proposed technique is integrated within the process of detecting and handling anomalies (that are potentially the result of a cyber-attack) in an operational, real-time, cyber-physical system.
The proposed approach also provides visual clues to assist the human observer in immediately understanding the cause of the alert and identifying false alarms, and enables improved performance of \methodname.

The proposed method highlights the most suspicious areas within an analyzed image.
The identification of these suspicious areas is performed as follows.
\begin{itemize}
    \item \textbf{Select the most inaccurately reconstructed image.} Given a window $w$ (i.e., a sequence of images) that is considered anomalous, select the image that was reconstructed in the most inaccurate way according to the SSIM score.
    We denote the original image by $in_i$ and the reconstructed image by $out_i$.
    
    \item \textbf{Divide the image to sub-images.} Split $in_i$ and $out_i$ into $n*n$ equal, corresponding, squares denoted by $in_{i,j}$ and $out_{i,j}$, where $j$ is between one and $n*n$.
    \item \textbf{Calculate the SSIM for sub-images.} Calculate the SSIM score for each of the corresponding pair of squares obtained from $in_{i,j}$ and $out_{i,j}$.
    \item \textbf{Highlight sub-images by their calculated SSIM score.} Assign a color to each $in_{i,j}$ and $out_{i,j}$ pairs based on the SSIM score calculated.
\end{itemize}

This method allows an experienced user to easily and quickly determine whether the result of the model is reliable and logical or whether it is a false alarm.

%% file: evaluation.tex
\section{Evaluation}

\subsection{Datasets}
For the evaluation, we used five datasets extracted from the OpenSky platform,\footnote{\url{https://opensky-network.org/. }} a participatory sensor network for air traffic research, containing real-world data on a large scale~\cite{strohmeier2015opensky}. 
OpenSky was chosen because it was the only free data source we found available that enabled us to obtain a vast amount of data from a certain geographic location.
Each dataset covers a 10,000 square kilometer area, centered around a different airport (see Table~\ref{tab:datasets}).

\begin{table}
\centering
\caption{\label{tab:datasets}Description of datasets.}
\begin{tabular}{ | C{1.1cm} | C{2.2cm} | C{2.2cm} | C{1.4cm} |} 
 \hline
 \textbf{Name} & \textbf{Location (center)} & \textbf{\# of messages} & \textbf{Duration}  \\ 
 \Xhline{3\arrayrulewidth}

 \rowcolor[gray]{0.9}
 Paris &  {CDG Airport} & 100,000 & {7 days} \\
  \hline
  
  Berlin &  {TXL Airport} & 800,000 & {7 days} \\
  \hline

\rowcolor[gray]{0.9}
 London &  {LHR Airport} & 1,000,000 & {7 days} \\
  \hline
   Rome &  {FCO Airport} & 700,000 & {7 days} \\
  \hline
  \rowcolor[gray]{0.9}
   Amsterdam &  {AMS airport} & 800,000 & {7 days} \\
  \hline
 
\end{tabular}
\end{table}

\subsection{Injected Anomalies}
In order to evaluate the performance of the trained model, we injected five types of anomalies into the flight data included in the test sets. 
Each attack was injected approximately 100 times for each dataset in the following way. 
For each 50 image segment in the test set we left the first 35 images unchanged, and in the remaining 15 images we injected an anomaly, thus enabling us to create an infected test set that includes a few sections containing a certain type of attack.
We used attacks that are reasonable for an attacker to implement, as described in Mirzaei \etal~\cite{mirzaei2019security}.
The attacks are as follows:

\textit{Flooding - } We inject a random number (between three and eight) of scattered aircraft at uniformly distributed random locations. 
In this scenario we are exploiting the lack of an authentication mechanism in order to make the ADS-B system unreliable because of the inability to distinguish between a real aircraft and a fictitious aircraft. 
An attack of this kind can occur without much effort by using COTS software as previously demonstrated ~\cite{schafer2013experimental}.

\textit{Ghost} - We inject an aircraft advancing along a legitimate route, but taken from a different geographical location (i.e., based on data that is not from the five datasets tested).
Previous research by Costin~\cite{costin2012ghost} and McCallie~\cite{mccallie2011security} discussed the dangers of impersonating (ghost) aircraft.

\textit{Jamming} - We make an aircraft disappear for a certain period of time. 
This scenario represents unexpected disappearance of an aircraft which can be caused by a denial-of-service attack, in which an aircraft is prevented from sending or receiving ADS-B messages, as demonstrated by Wilhelm \etal~\cite{wilhelm2011short}.

\textit{Reverse} - We change the direction of the arrow by 180 degrees for a certain period of time, thus creating an abnormal movement pattern.

\textit{Change altitude} - We change the altitude of the aircraft by first checking whether the altitude of the aircraft is above or below a certain threshold (in order to determine whether the aircraft is descending for landing or not), and then we switch the altitude accordingly: if the aircraft is descending we change the value to a high altitude, and vice versa.

\begin{figure}
\centering
\includegraphics[width=9cm,keepaspectratio]{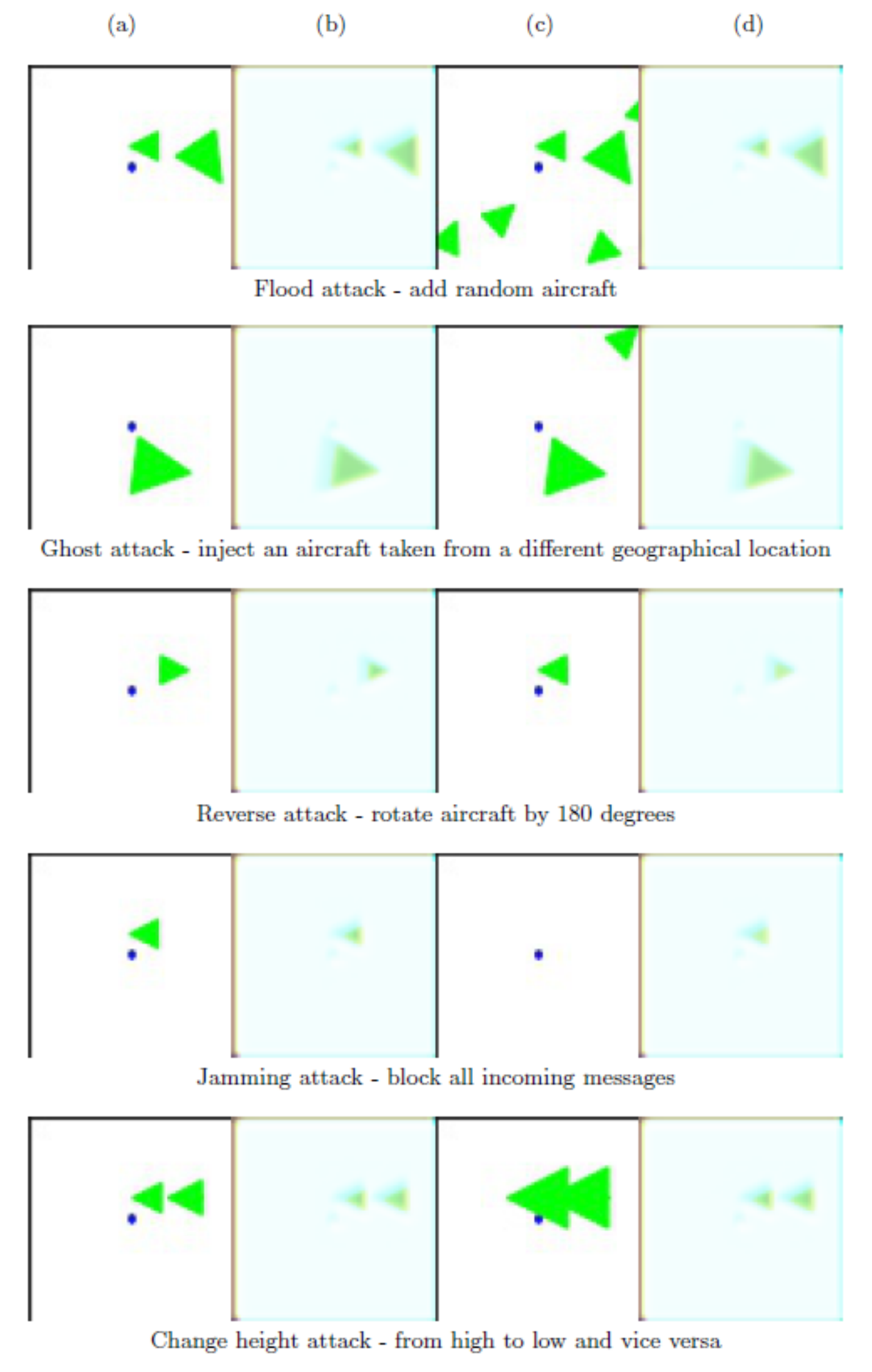}

\caption{Illustration of the five attacks injected and evaluated in the experiments. 
Each row represents a different injected attack. 
The columns present the original image before injecting the attack (a), the image reconstructed by applying the model on the original image (b), the image after injecting the attack (c), and finally, the image reconstructed by applying the model on the image after injecting the attack(d).}
\label{fig:attacks-example}
\end{figure}

An illustration of the five injected attacks is presented in Figure~\ref{fig:attacks-example}. 
Each row represents a different injected attack.
The columns represent the following: (a) original image before injecting the attack, (b) the image reconstructed by applying the model on the original image, (c) the image after injecting the attack, and (d) the image reconstructed by applying the model on the image after injecting the attack. 
It can be seen that the difference between the input images (the original image (a) and the image after injecting the attack (c)) and the reconstructed images is much higher for the images with the injected attack, demonstrating the ability to detect the attacks.

\begin{figure}
\centering
\includegraphics[width=8cm,keepaspectratio]{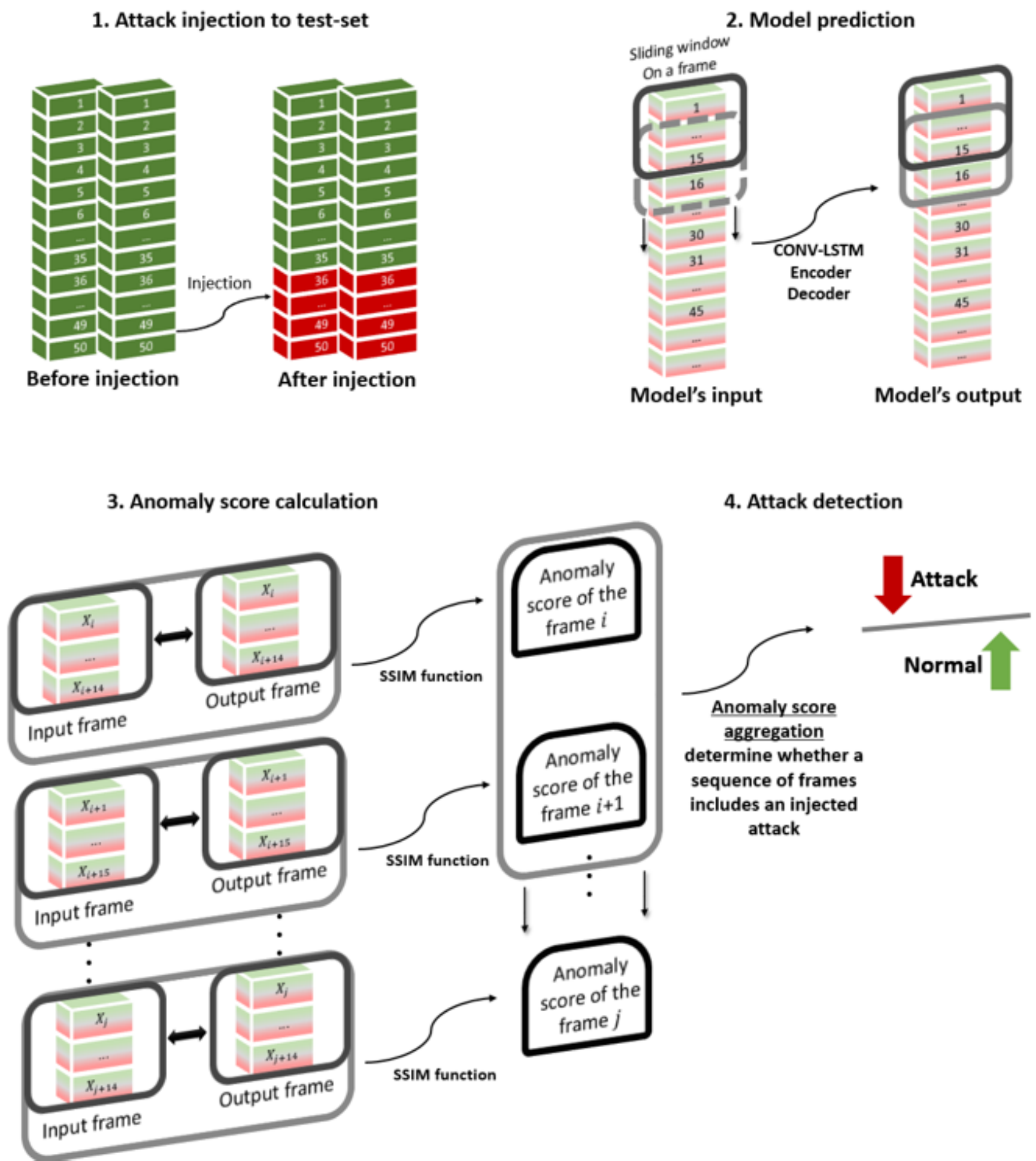}
\caption{\label{fig:process}Representation of the process of injecting anomalies and calculating the anomaly score. 
In the first stage the attacks are injected into the dataset; in the second stage, the model processes the dataset and creates the output frames accordingly.
Then ,in the third stage, the SSIM score for each frame is calculated. 
Finally, in the fourth stage, aggregation is carried out according to the threshold values $t_1$ and $t_2$. }
\end{figure}

\subsection{Evaluation Process}
\subsubsection{Anomaly Detection Evaluation}
We divided each dataset into three subsets: training, validation, and test sets. 
The training set was used for training the model, the validation set was used for setting the threshold value, and the test set was used for evaluating the model. 
The training set consisted of approximately 60\% of each dataset, the validation set consisted 25\% of each dataset, and the test set consisted of the remaining 15\%.

We set the parameters as follows:
\begin{itemize}
    \item $s$ = 15 images;
    \item $t_1$ is the fifth percentile of the SSIM values from the validation set;
    \item several $\Delta t$ values were tested - $\Delta t$ = 2, 5, 10, and 20s with a 50\% overlap between consecutive images; and
    \item $t_2$ = 5.
\end{itemize}

We then applied the detection model on the test set and the SSIM value for the frames was calculated. 
The frames that exceed $t_1$ were aggregated. 
If the number exceeded $t_2$, we considered the window anomalous.
The overall evaluation process is shown in Figure~\ref{fig:process}.

\subsubsection{XAI Evaluation} 
To evaluate the ability of the proposed XAI technique in assisting the pilot in interpreting alerts and identifying false alarms, we present the output of the XAI module, i.e., the original and reconstructed images marked with the most anomalous areas to a human domain expert and checked whether the human domain expert was able to identify actual attacks and false positives.

\subsection{Results}
\begin{figure}
  \centering
  \includegraphics[scale=0.5]{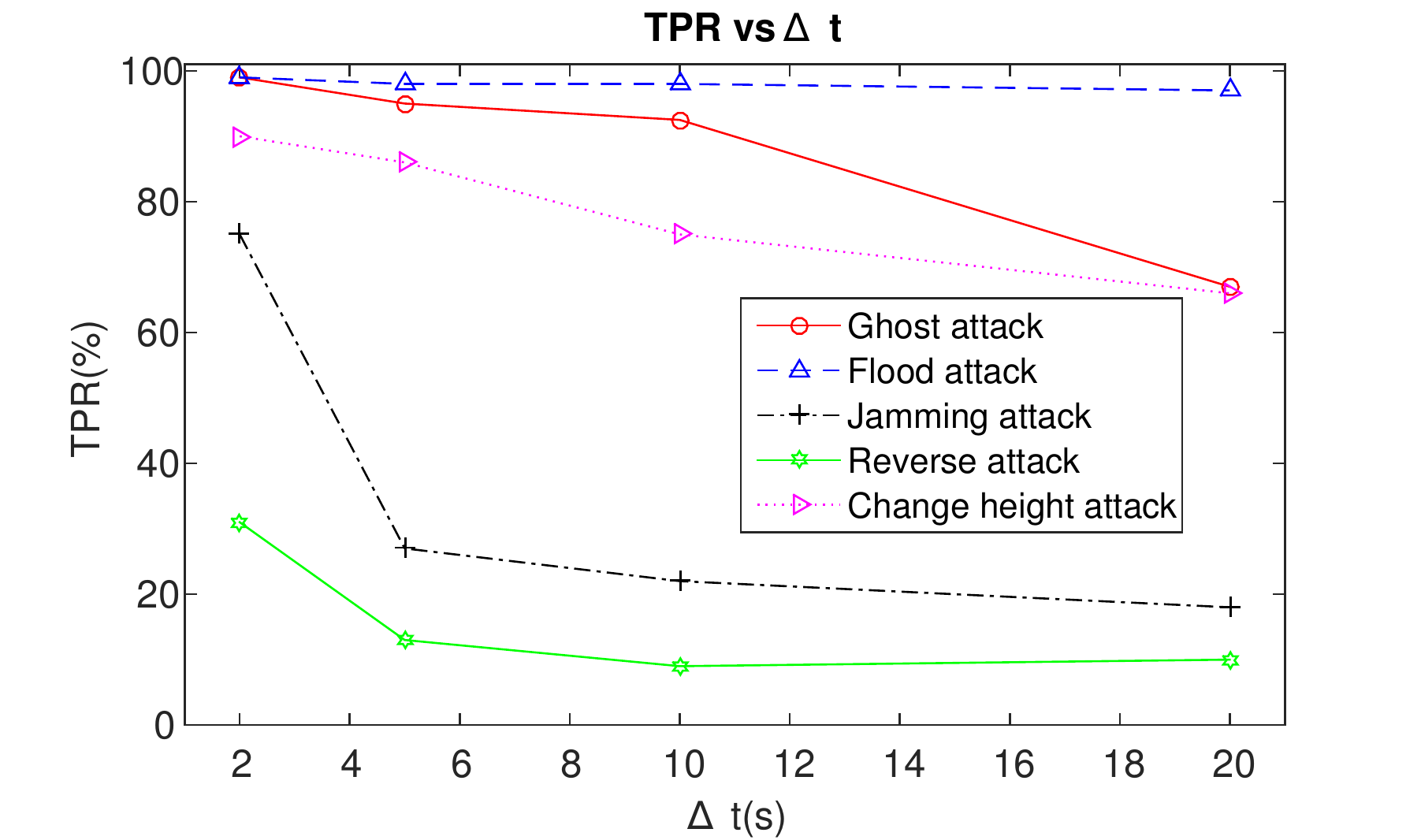}
  \caption{TPR values of the London dataset for different values of $\Delta$ t. 
  it is obvious from this figure that in general as $\Delta$ t grows bigger the TPR in all attacks goes worse.}
\label{dt2}
\end{figure}

\begin{figure}
  \centering
  \includegraphics[scale=0.5]{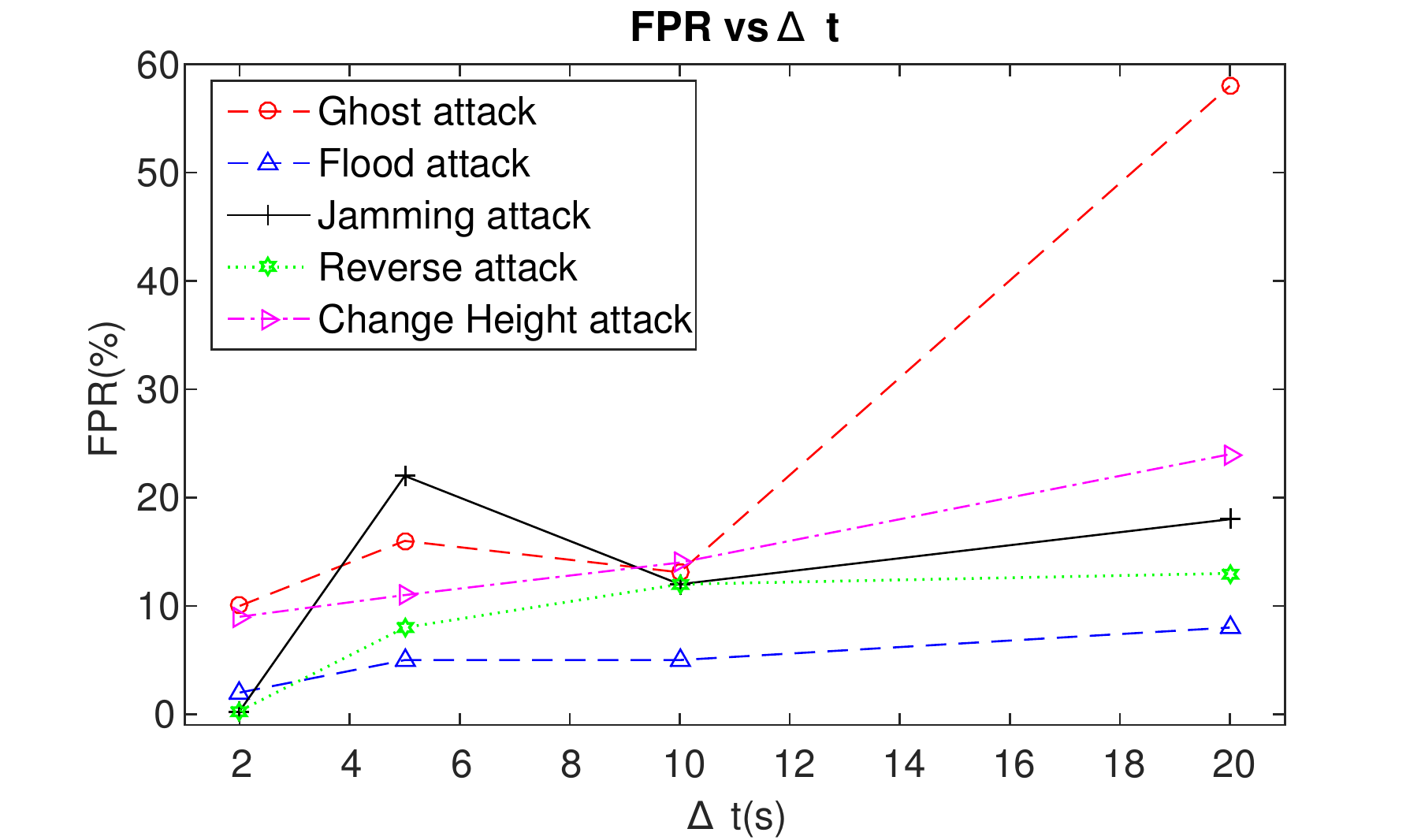}
  \caption{FPR values of the London dataset for different $\Delta$ ts. it is obvious from this figure that in general as $\Delta$ t grows bigger the FPR in all attacks goes worse.}
\label{dt1}

\end{figure}

Out of the tested $\Delta t$ values, on several of the datasets examined, it seems that for the length of the attacks tested (15 seconds), the best performance was achieved for $\Delta t$ = 2s.
$\Delta t$ = 2s means that the number of images is equal to the number of seconds - as the images overlap. 
The TPR and FPR of the attacks for the London dataset can be seen in Figures~\ref{dt1} and \ref{dt2}.
We can also see that in general, the lower $\Delta t$ is, the better the performance - for this attack length.
This observation is reasonable, since if $\Delta t$ is longer than the length of the attack, it will be harder for the model to capture the attack, since the attack spreads over fewer images.


\begin{figure*}
\centering
\includegraphics[width=12cm,keepaspectratio]{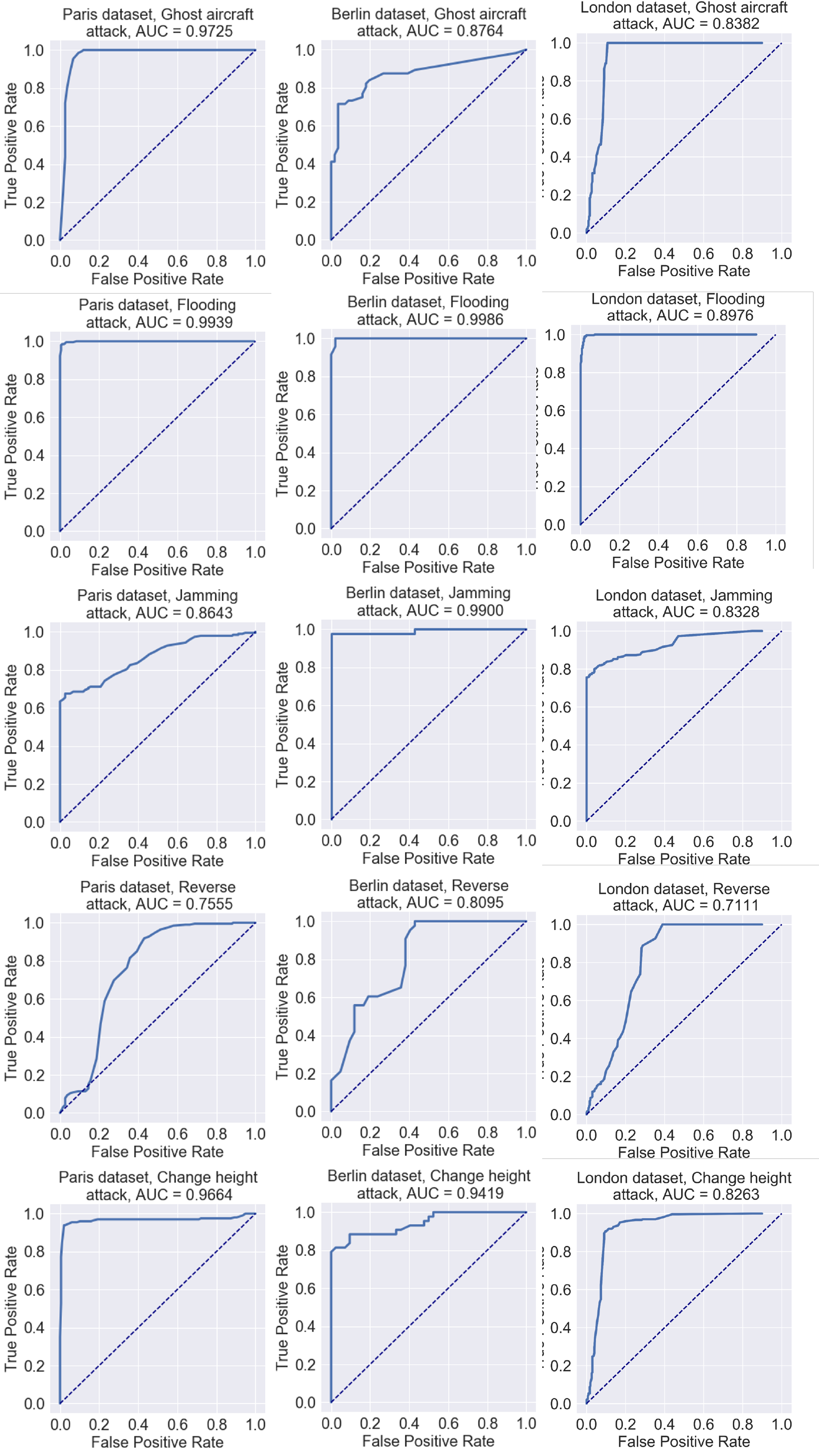}
\caption{ROC curves and matching AUC values for each dataset and attack.}
\label{fig:all_roc}
\end{figure*}

\begin{figure}
\centering
\includegraphics[width = 8cm,keepaspectratio]{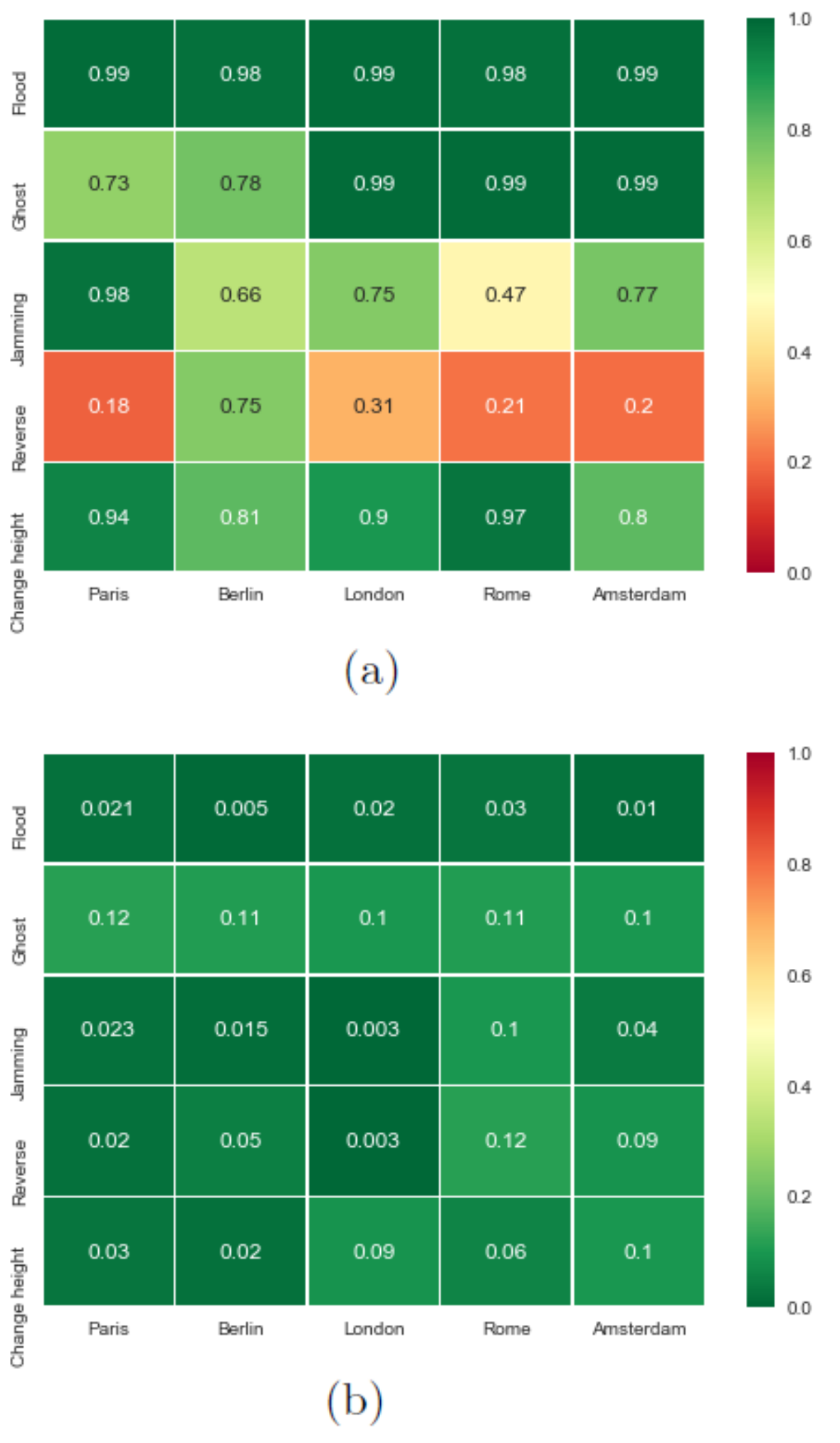}
\caption{(a) TPR and (b) FPR of different attacks and databases for $\Delta t = 2s$.}
\label{TPRFPR}
\end{figure}

Figure~\ref{fig:all_roc} presents the ROC curves for the datasets and attacks, and Figure~\ref{TPRFPR} presents the false positive rate (FPR) and true positive rate (TPR) values derived from the ROC curves for the selected thresholds.
The SSIM threshold was chosen separately for each dataset, based on the validation set and the desired TPR as opposed to the desired FPR.
We believe that choosing the SSIM threshold for each dataset is legitimate because of the differences between the different geographical locations.
Our method permits a slightly higher FPR, because, as discussed below, the XAI technique we can deal with it.

It can be seen (in Figure~\ref{TPRFPR}) that the model was able to detect the flood, ghost, and change altitude attacks on most datasets, and the Jamming attack was detected in most cases. 
It is reasonable that the attacks that involve adding aircraft are detected better than attacks that involve altering the data of existing aircraft.

The TPR obtained for the attacks indicates that meaningful features of the legitimate data were captured by the hidden state of the model.
For example, for the ghost attack the model was able to learn legitimate flight routes and identify unusual routes.
The detection rate of the ghost attack was the highest for the London dataset.
We attribute that result to the size of the dataset (Heathrow airport is rated as the seventh busiest airport in the world based on the number of flights~\footnote{\url{https://en.wikipedia.org/wiki/List_of_busiest_airports_by_passenger_traffic\#Preliminary_2018_statistics}}) which provided the model with sufficient data to capture legitimate flight routes.

The results show that the reverse attack is hard to detect (by the low TPR).
This can be explained by the fact that the reverse attack affects a smaller area in the image, and therefore is more subtle and difficult to detect; these changes may be better captured by an attention based CNN~\cite{xu2015show} which provides the ability to focus the model's attention on a subset of its inputs (i.e., specific regions within the image).

It can also be noted that the results for each attack vary a bit between datasets.
Based on our understanding, it can be explained by the fact that normal behavior is different in different geographical areas, for example, if at a certain airport there are a lot of aircraft that normally fly nearby at a high altitude, the change altitude attack would not be detected as well as it would be at other airports (and this might explain the Amsterdam or Berlin datasets result in this attack).
Another example is that in an airport with several distinct flight routes approaching it, the ghost attack might be better detected than at an airport that is accessed from all directions (thus making the detection of the anomaly more difficult).

The results obtained by our method are comparable to results obtained in a previous study, that dealt with single routes of aircraft~\cite{habler2018using}.

\subsubsection{XAI Results}
\begin{figure}
\centering
\includegraphics[width=5cm,keepaspectratio]{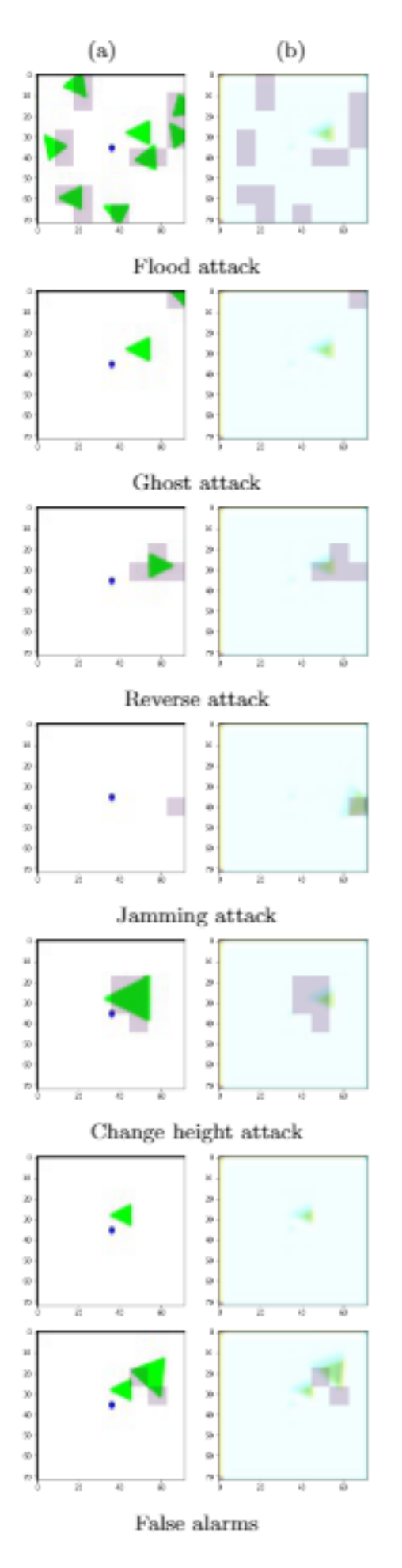}
\caption{several suspicious frames, explained by our technique. (a) is the input to the model and (b) is the output. It can be seen that the interesting parts for the observer to examine are the parts in purple, and that the anomalies are near those areas, while the parts in yellow represents the areas without anomalies. Using this technique allows the observer to identify false alarms easily.}
\label{fig:XAI_exapmles}
\end{figure}

Using the XAI technique, we were able to call attention to "suspicious" frames that has been detected as anomalous. 
The observer was then able to identify all false alarms within an average of five seconds for each frame. 
Thus, without affecting the model's TPR rate, we managed to obtain the observer's trust by providing a visual indicator of the model's result, while maintaining an inspection rate of less than ten suspicious frames per hour.
An example of true attacks and false alarms and their representations can be seen in Figure~\ref{fig:XAI_exapmles}.

%% file: conclusions.tex
\section{Conclusion and Future Work}

In this study, we proposed a method for detecting anomalous ADS-B messages in a specific geographical area; the method, inspects all aircraft within the chosen airspace at once and is based on applying a ConvLSTM model on the data.
By using known methods for image analysis and representing the information within a geographical area as images, we were able to represent information from all of the aircraft located in the examined area without the need of feature engineering.
We were also able to analyze the input and output using the XAI technique, enabling a human observer differentiate false alarms and actual detections.

Our method enables the detection of attacks generated by sophisticated attackers that can overcome other security measures.

We validated our model using several geographical locations and five types of injected attacks, obtaining satisfactory TPR and FPR values for most of those attacks.

Based on the experiments conducted, we can conclude that it is possible to assess the integrity of messages from the ADS-B protocol without changes to the protocol or its architecture. 

In future work, we plan to test the model using different and more subtle attacks and additional locations, while tuning the hyper-parameters of the model and its architecture. 
Another direction for future research is to use the work of Habler \etal~\cite{habler2018using} in order to enhance the performance of our model by using different colors for suspicious aircraft (e.g., legitimate ADS-B messages in green and suspicious message in red). 
Another research direction revealed during this research is to combine several networks with different $\Delta t$ values in order successfully deal with attacks of several lengths.
We also plan to evaluate the robustness of the model against adversarial attacks.